# Dynamics of a stochastically driven running sandpile

Torsten Becker, Heiko de Vries and Bruno Eckhardt

*Institut für Chemie und Biologie des Meeres and*

*Fachbereich Physik der C.v. Ossietzky Universität,*

*Postfach 25 03, D-26111 Oldenburg, Germany*

May 30, 1994

## Abstract

We analyze in detail a one-dimensional stochastically driven running sandpile. The dynamics shows three different phases, depending on the on-site relaxation rate and stochastic driving rate. Two phases are characterized by the presence of travelling waves. The third shows algebraic relaxation.

# 1 Introduction

The investigation of a simple automaton with nontrivial time correlations by Bak, Tang and Wiesenfeld [1] has triggered a number of studies on a variety of similar models. Whereas the initial model was deterministically driven at a vanishing rate (every perturbation was allowed to relax to a stationary state before a new perturbation was added), later models (for instance [3, 4]) allowed for stochastic driving at a finite rate. The case of infinitely extended lattices has been analyzed using methods from renormalization and mode coupling theory ([4]). For finite lattices, multifractal scaling has been used to analyze finite size effects ([2]).

We here analyze for a specific automaton both the deterministic dynamics and the stochastic relaxation in some detail. Although there is no system which is exactly described by this automaton, it helps to have in mind some physical processes with qualitatively similar behaviour. As in the Bak, Tang, Wiesenfeld [1] model, one may interpret our automaton as an idealized one-dimensional sand pile. Because of the boundary conditions, it is more like a sand pile near a wall. Sand is added randomly at a site, which topples if the slope exceeds a certain threshold. Alternatively, one may think of a chain of water reservoirs, where water is added randomly and where the reservoirs spill over, if a certain water level is exceeded. We assume that the amount of mass transported per time step in the case of a supercritical slope consists of a constant part plus an additional slope dependant part. One expects and finds in numerical simulations that, after some transients, the automaton relaxes to a statistically stationary state with nontrivial dynamics.

Analysis of this stationary state reveals three different kinds of phases, depending on the driving rate and the relaxation rate. In the Bak, Tang, Wiesenfeld [1] case of infinitesimal driving rate the dynamics is rather trivial and consists of waves which pass through the system with rather little interaction. As the driving rate is increased, there is first a regime dominated by a different kind of waves, and then a regime with algebraic relaxation because of interaction between relaxation and driving. The boundaries between the different regimes can be determined and are in agreement with numerical simulations.

In section 2 we define the model and its parameters and present simulation results for the slope profile of the statistically stationary state, for the dynamics of perturbations and for the temporal correlations at a fixed site.

In section 3 we discuss some idealized solutions to the dynamical rules which may be thought of as waves and wave packets. In some regions of the lattice they exist as undamped running waves whereas in others they are exponentially damped. A transport argument allows to explain quantitatively the structure of the slope profiles.

In section 4 we use methods from stochastic calculus to obtain a formally exact solution for the relaxation dynamics. This allows us to discuss not only the low frequency behaviour which is not accessible in direct numerical calculations, but also to estimate the transition between the various scaling regimes in the power spectra.

In section 5 we compare our results with previous ones and comment on their relevance to other, in particular higher dimensional automata.

## 2 Definitions and simulation results

### 2.1 The model

As described above, one may think of our model as a chain of water reservoirs or as a one-dimensional sand pile. This picture suggests to use the water level at each reservoir or the height of the sandpile at each site as variables. However, the dynamics is controlled not so much by the absolute level but rather by the amount of mass transported, so that the slopes, i.e. differences between neighboring sites, are better suited variables for these automata.

The state of the system may thus be described by a vector $\vec{h} = (h_0, ..., h_{N-1}) \in R^N$ for the heights or by a vector $\vec{\sigma} = (\sigma_0, ..., \sigma_{N-1}) \in R^N$ for the slopes. The interpretation of these vectors can be seen in Fig. 1. The transform

$$\vec{h} = (h_0, ..., h_{N-1}) \longmapsto \vec{\sigma}(\vec{h}) = (\sigma_0, ..., \sigma_{N-1}) = (h_0 - h_1, ..., h_{N-2} - h_{N-1}, h_{N-1}) \quad (1)$$

maps between the two description pictures. Henceforth we will use the slope picture. The dynamics consists of two steps: A relaxation rule and a perturbation rule (the addition of mass). The state $\vec{\sigma}$ at time $t$ evolves into the state $\vec{\sigma}'$ at time $t+1$ according to these steps:

**(i)** Relaxation rule

If a slope $\sigma_i$ is too steep it will relax by a certain amount $r_i$ of mass being transferred to the right neighbouring site. $r_i$ is given by

$$r_i := \begin{cases} 1 + (\sigma_i - \sigma_c)/P & \text{if } \sigma_i > \sigma_c \\ 0 & \text{else} \end{cases} \quad , i = 0, \ldots, N-1. \quad (2)$$

One arrives at an intermediate state $\sigma^{\text{tmp}}$:

$$\left. \begin{array}{rcl} \sigma_0^{\text{tmp}} & = & \sigma_0 - 2r_0 + r_1 \\ \sigma_i^{\text{tmp}} & = & \sigma_i + r_{i-1} - 2r_i + r_{i+1} \, , \, i = 1, \ldots, N-2 \\ \sigma_{N-1}^{\text{tmp}} & = & \sigma_{N-1} + r_{N-2} - r_{N-1} \end{array} \right\} \quad (3)$$

The relaxation rule contains two parameters, a critical slope $\sigma_c$ and a relaxation rate $P$.

**(ii) Perturbation rule**

A site $i_0$ of the lattice is chosen randomly and a mass $s \in R^+$ is added to that site. For the first site, $i_0 = 0$, we have

$$\sigma'_i := \begin{cases} \sigma_i^{\text{tmp}} & , i \neq 0 \\ \sigma_i^{\text{tmp}} + s & , i = 0 \end{cases}, \quad i = 0, \ldots, N-1 \quad (4a)$$

and for all other sites, $i_0 \neq 0$,

$$\sigma'_i := \begin{cases} \sigma_i^{\text{tmp}} & , i \neq i_0, i_0 - 1 \\ \sigma_i^{\text{tmp}} + s & , i = i_0 \\ \sigma_i^{\text{tmp}} - s & , i = i_0 - 1 \end{cases}, \quad i = 0, \ldots, N-1 . \quad (4b)$$

Here another parameter $s > 0$ characterizing the driving rate enters. The perturbation rule is illustrated in Fig. 1a and 1b.

We expect the model to evolve from an arbitrary initial state into a statistically stationary state. In order to recognize the time when the model reaches this state we averaged the slope profiles in blocks of $100N$ time steps and calculated the relative differences in the profiles between consecutive blocks. If this relative error became less then a prescribed value we assumed the model to have reached the stationary state. To be on the safe side averages were computed after an additional iteration of 10 times the number of steps needed so far. The value of the critical slope $\sigma_c$ only affects the average slope, i.e. the height of the stationary state in the slope picture, but it has no effect on the structure of the stationary state or dynamical properties. The essential parameters are therefore $P$ and $s$; for the trivial parameter $\sigma_c$ we chose $\sigma_c = 4$ throughout the paper. In the next two subsections we illustrate the properties of the stationary state.

## 2.2  Slope profiles in the stationary state

The model relaxes to different stationary states depending on the two parameters $P$ and $s$. Fig. 2, a division of the $P$-$s$-plane which separates automatons with structurally different stationary states, directs attention to Fig. 3-7 in which time averaged slope profiles along with some snapshots out of certain lattice parts are shown. Common to these figures is the choice for the lattice size $N = 200$ and the critical slope $\sigma_c$ (the critical slope is represented by dashed lines in the figures). We have omitted plotting the average slope at the last lattice site in each profile since it is affected by a trivial boundary effect.

One can infer from these figures the apparent importance of wave packets which are visible in all of the snapshots but frame 3 in Fig. 7. We can distinguish packets with averages below and above the critical slope $\sigma_c$. Following the time evolution of these profiles one notes that the wave packets are stable and propagate one site per time step. If a wave packet is perturbed by the addition of mass the resulting propagating profile is still a wave packet but it can contain a different number of periods than before. Wave packets arise from perturbations in parts of the lattice with constant slope. The exceptional case mentioned above (part 3 of the profile for $P = 6$, $s = 2$) exhibits another kind of behaviour: The addition of mass results in a perturbation of the straight line which will be damped during the time evolution.

We refer to different parts of the stationary state as phases characterized by the excitations they carry: Subcritical wave packets in subcritical regions, supercritical wave packets in supercritical regions and damped perturbations in linearly increasing lattice parts. The division of the $P$-$s$-plane in Fig. 2 separates automatons with different numbers of phases: The ones in region I contain only the phase with subcritical waves, the ones in region II contain the first two phases and the ones in region III all three phases.

## 2.3 Temporal correlations

Temporal correlations in regions of linearly increasing average slope (phase 3) show algebraic decay as demonstrated in Fig. 8 for the parameters $P = 10$, $s = 2$. Fig. 8a shows the autocovariance function of the slope at a fixed position in this phase. The spectral density function plotted in Fig. 8b consists of two clearly distinguishable regions of algebraic behaviour with different slopes. Increasing the value of $P$ leads to higher scaling exponents.

In the next two sections we attempt to explain the simulation results.

# 3 Wave packets and transport

Wave packet like structures are characteristic excitations in the first two phases. As will be shown in subsection 3.1 the dynamical rule of the model supports certain exact running waves and wave packet like solutions which are undamped. In contrast, excitations in the third phase are damped as will be proved in subsection 3.2 thus explaining the absence of wave packets in linearly increasing regions of the lattice. A transport argument given in subsection 3.3 based on these excitations finally explains the partition of the parameter plane.

## 3.1 Idealized waves and wave packets

We consider in this subsection a simplified system with an infinite number of sites, i.e. the state space of the system is the set of series $(\sigma_n)_{n \in Z}$. The dynamic reduces to the relaxation rule

$$r_i := \begin{cases} 1 + \frac{\sigma_i - \sigma_c}{P} & , \sigma_i > \sigma_c \\ 0 & , \text{else} \end{cases} \quad (5)$$

$$\sigma'_i = \sigma_i + r_{i-1} - 2r_i + r_{i+1} , \quad \text{for all } i \in Z . \quad (6)$$

An idealized infinitely extended wave is a state

$$\left( \begin{cases} c + d & , n \text{ even} \\ c - d & , n \text{ odd} \end{cases} \right)_{n \in Z} , \quad (7)$$

characterized by two parameters $c \in R$ and $d > 0$, whose evolution in time is simply a shift of one place per time step. A subcritical wave packet is one with an average below criticality, i.e. $c \leq \sigma_c$ and $c + d > \sigma_c$; a supercritical wave packet is one with an average above criticality, i.e. $c > \sigma_c$ and $c - d \leq \sigma_c$. The pairs $(c, d) \in R \times ]0, \infty[$ for which an idealized wave obeys the dynamical rule in eq. (6) obviously depend on the parameter $P$ through eq. (5). One finds for

- subcritical waves

$$\left. \begin{aligned} P > 1 \quad &: \quad W_{\text{sub}} = \{(c, d) \mid \sigma_c - 1 < c \leq \sigma_c , \\ & \qquad\qquad d = 1 + \tfrac{c+1-\sigma_c}{P-1} \} \\ P = 1 \quad &: \quad W_{\text{sub}} = \{(c, d) \mid c = \sigma_c - 1 , d > 1\} \\ 0 < P < 1 \quad &: \quad W_{\text{sub}} = \{(c, d) \mid c < \sigma_c - 1 , \\ & \qquad\qquad d = 1 + \tfrac{c+1-\sigma_c}{P-1} \} \end{aligned} \right\} \quad (8)$$

- and for supercritical waves

$$\left. \begin{aligned} P > 2 \quad &: \quad W_{\text{super}} = \{(c, d) \mid \sigma_c < c \leq \sigma_c + \tfrac{P}{P-2} , \\ & \qquad\qquad d = 1 + \tfrac{c+1-\sigma_c}{P-1} \} \\ 1 < P \leq 2 \quad &: \quad W_{\text{super}} = \{(c, d) \mid \sigma_c < c , d = 1 + \tfrac{c+1-\sigma_c}{P-1} \} \\ 0 < P \leq 1 \quad &: \quad W_{\text{super}} = \emptyset . \end{aligned} \right\} \quad (9)$$

In the subcritical case it is easy to see that besides these infinite structures wave packets with finite extent are also solutions of the dynamical rules. A wave packet of finite extent has a certain number of periods $(c-d, c+d)$ or $(c+d, c-d)$, bracketed by infinite sequences of $c$, for instance

$$\sigma = (\ldots, c, (c-d, c+d)_n, c, \ldots) . \tag{10}$$

This structure (10) propagates one site to the right per time step.

Finite wave packets are results of perturbations in constant parts of the lattice: It is for instance possible to observe in the time development of the state

$$\sigma = (\ldots, c, c-s, c+s, c, \ldots) , \ c \leq \sigma_c , \ c+s > \sigma_c , \tag{11}$$

obtained by adding mass to a certain site, the formation of a finite wave packet where the number of periods depends on the choice of the parameters $P, s$ and the mean value $c$.

In the supercritical case things are a little more complicated: There exists no state with only a finite number of components deviating from a constant $c$ which propagates without changing its form. But it is possible to construct propagating structures which are similar to ideal wave packets of finite extent. Consider for instance a state with the properties

$$\left. \begin{array}{c} \forall i \leq 0 \ : \ \sigma_i = c > \sigma_c \\ \\ \sigma_1 < \sigma_c \\ \\ \sigma_2 > \sigma_c \\ \\ \sigma_3 < \sigma_c \\ \\ \forall i \geq 4 \ : \ \sigma_i > \sigma_c \end{array} \right\} \tag{12}$$

A necessary and sufficient condition for this state to propagate backwards one site per time step without changing its form requires special values for the components $\sigma_i$:

$$\left. \begin{array}{rcl} \sigma_1 & = & c - 1 - \frac{c-\sigma_c}{P} \\ \\ \sigma_2 & = & \frac{P(c+1)-\sigma_c}{P-1} \\ \\ \sigma_3 & = & \sigma_2\left(1-\frac{2}{P}\right) - 2 + 2\frac{\sigma_c}{P} \\ \\ \sigma_4 & = & \frac{P\sigma_3+\sigma_2+2(P-\sigma_c)}{P-1} \\ \\ \sigma_5 & = & \frac{(P-2)\sigma_4+\sigma_c-P}{P-1} \\ \\ \sigma_n & = & \frac{1}{P}\sigma_4 + \frac{P-1}{P}\sigma_5 + \frac{1-P}{P}(\sigma_5 - \sigma_4)\frac{(-1)^{n-4}}{(P-1)^{n-4}} \\ \\ & & (n \geq 6) \end{array} \right\} \tag{13}$$

The possibility of wave packets with an arbitrary number of periods in the sub- and the supercritical case is of great importance for the transport argument below.

## 3.2 Damping of perturbations

We consider a part of the lattice with always supercritical slope. Then the relaxation rule simplifies to

$$r_i = 1 + \frac{\sigma_i - \sigma_c}{P} = 1 - \frac{\sigma_c}{P} + \frac{\sigma_i}{P} \tag{14}$$

$$\sigma'_i = \frac{1}{P}\sigma_{i-1} + \frac{P-2}{P}\sigma_i + \frac{1}{P}\sigma_{i+1} \ . \tag{15}$$

We are particularly interested in states which depend linearly on the index $i$ because the time averaged slope profile in part 3 of the lattice shown in Fig. 5 is of that form. One property of the relaxation rule (15) is that it leaves straight lines $\sigma_i = c + ib$ invariant. Localized perturbations of these straight lines expand only one site per time step and thus remain localized. For the evolution of such perturbations it then suffices to study the dynamics on a finite sublattice of size $N$. Because of linearity of the dynamics described by (15) we may separate the evolution of the perturbation $\vec{\varepsilon}(n) \in R^N$ ($n$ is the time step) from an underlying time independent straight line and write down a vector equation

$$\vec{\varepsilon}(n+1) = M\vec{\varepsilon}(n) = M^{n+1}\vec{\varepsilon}(0) \tag{16}$$

where

$$M = \begin{pmatrix} \frac{P-2}{P} & \frac{1}{P} & 0 & \cdots & 0 \\ \frac{1}{P} & \ddots & \ddots & \ddots & \vdots \\ 0 & \ddots & \ddots & \ddots & 0 \\ \vdots & \ddots & \ddots & \ddots & \frac{1}{P} \\ 0 & \cdots & 0 & \frac{1}{P} & \frac{P-2}{P} \end{pmatrix} \quad (N \times N - \text{matrix}) \tag{17}$$

is the relaxation matrix. Since $M$ is symmetric it may be diagonalized. Its eigenvalues are

$$\mu_k = \frac{P-2}{P} - \frac{2}{P}\cos\frac{k\pi}{N+1} \ , \quad k = 1, \ldots, N \tag{18}$$

with the corresponding eigenvectors

$$\hat{v}_n^{(k)} = (-1)^{n+1}\sqrt{\frac{2}{N+1}}\sin\frac{nk\pi}{N+1} \ , \quad n = 1, \ldots, N \ . \tag{19}$$

After that diagonalization the derivation of a formal expression for $\vec{\varepsilon}(n)$ leads to

$$\varepsilon_i(k) = \sum_{\eta,\kappa=1}^{N} \mu_\kappa^k \hat{v}_i^\kappa \hat{v}_\eta^\kappa \varepsilon_\eta(0) \ , \quad i = 1, \ldots, N \ . \tag{20}$$

Since for $P > 2$ all eigenvalues have an absolute value less than one, any perturbation will be damped. For the model this means that in parts of the lattice which are always supercritical all deviations from a straight line tend to vanish during time evolution.

## 3.3 Transport argument

The mean transport rate at site $i$ of the lattice must be $(i+1)\frac{s}{N}$ since the average mass added per time step at a fixed site is $s/N$. In order to handle this flux time evolutions with the right amount of excess slope are necessary.

Consider first the case $P > 2$. Subcritical wave packets separated by regions of slope less than $\sigma_c$ cover a transport rate range of $\left[0, \frac{1}{2}\frac{P}{P-1}\right]$ since the transport in regions of undercritical slope vanishes, while in a region containing a wave packet with average $c = \sigma_c$ it is maximal: Even sites contribute $r_i = \frac{P}{P-1}$ and odd sites nothing, resulting in an average transport rate $\frac{1}{2}\frac{P}{P-1}$. From these considerations we conclude that for $s < \frac{1}{2}\frac{P}{P-1}$ the transport can be achieved through subcritical wave packets. An important point is the variability in the length of the wave packets and in the subcritical regions between them which leads to a variety of time evolutions covering the whole transport rate range $\left[0, \frac{1}{2}\frac{P}{P-1}\right]$. But all of these evolutions have the same time averaged slope profile: A horizontal subcritical line. This explains the plateaus seen in Fig. 3.

For $s > \frac{1}{2}\frac{P}{P-1}$ time evolutions with transport rates greater than the rates of subcritical waves must enter the dynamics. Larger rates can be achieved by supercritical wave packets separated by parts of constant supercritical slope. The minimum rate is attained by a packet that extends over the entire lattice region considered since now only every second site is supercritical. Its transport rate is exactly $\frac{1}{2}\frac{P}{P-1}$. The state with the maximum rate has a region with constant supercritical slope. The maximum slope in such a region is $\sigma_c + \frac{P}{P-2}$ because we still want to retain the possibility of supercritical wave packets. Thus transport rates in $\left[\frac{1}{2}\frac{P}{P-1}, \frac{P-1}{P-2}\right]$ can be achieved by states consisting of supercritical waves separated by regions of constant supercritical slope.

Therefore we expect in the case $P > 2$ and $\frac{1}{2}\frac{P}{P-1} < s < \frac{P-1}{P-2}$ a stationary state consisting of two regions: A constant subcritical region proceeding from the left boundary and a constant supercritical region extending to the right boundary. The transition occur approximately at the site at which the needed transport rate exceeds the maximum rate that can be achieved by subcritical waves. From this one estimates that the transition between region 1 and 2 should occur near

$$n_{1\to 2} = \frac{1}{2}\frac{P}{P-1}\frac{N}{s} - 1 \ . \tag{21}$$

These considerations are also in agreement with the simulation results as can be seen in Fig. 4.

At parameter values $s > \frac{P-1}{P-2}$ supercritical waves do not suffice to provide the necessary transport. This implies the existence of a third region in the stationary state which is different from the first two regions. The excess slope is so large that no wave packets are possible due to the damping of perturbations in this part. Therefore, there can be no variability in transport rates with time evolutions with the same time averaged horizontal profile. We expect in this case an increasing slope in order to adjust the transport. The transition point between second and third region can be estimated by an argument analogous to the one utilized above to be

$$n_{2 \to 3} = \frac{P-1}{P-2} \frac{N}{s} - 1 \ . \tag{22}$$

Furthermore, one can show that the slope of the profile in the third region must be $P\frac{s}{N}$ and the averaged value at site $N-1$ must be $<\sigma_{N-1}> = P(s-1) + \sigma_c$ (because of flux continuity). All these predictions are consistent with simulation results (cf. Fig. 5).

For the case $P \in ]1, 2[$ analogous considerations apply. Subcritical waves cover a transport rate range $\left[0, \frac{1}{2}\frac{P}{P-1}\right]$ while every rate larger than $\frac{1}{2}\frac{P}{P-1}$ can be achieved by supercritical wave packets. For $s < \frac{1}{2}\frac{P}{P-1}$ we expect the stationary state to be a subcritical line which is confirmed by the results depicted in Fig. 4. For $s > \frac{1}{2}\frac{P}{P-1}$ a region with time averaged slope greater than $\sigma_c$ is necessary. In this region supercritical packets are characteristic of the dynamics. An estimation of the site at which the profile crosses the critical value is again

$$n_{1 \to 2} = \frac{1}{2}\frac{P}{P-1}\frac{N}{s} - 1 \ . \tag{23}$$

Since for these values of the transport parameter $P$ there is no limitation for the mean value $c$ all transport with high rate can be done by supercritical wave packets. But in order to adjust the transport rate to increasing (with site index $i$) demands we expect the mean value $c$ to increase also with the site index $i$. These estimates are in good agreement with simulations (cf. Fig. 7).

The arguments given above finally lead to a partition of the $P$-$s$-plane in three distinct parts separated by the curves $s = \frac{1}{2}\frac{P}{P-1}$ and $s = \frac{P-1}{P-2}$ as shown in Fig. 2.

# 4 Calculation of temporal correlations

In this section we concentrate on the third region of the stationary state for the parameter choice $P > 2$, $s > \frac{P-1}{P-2}$. Our aim is to derive a closed form of the time correlations of the slope at a fixed site in this region. The derivation is analogous to the one given in [5] for stochastically driven ordinary differential equations. Since in this region the slope is supercritical all the time the relaxation rule takes on a simple form. We simplify further by adjusting the left and right boundary conditions. Together with the perturbation the simplified system can be regarded as a vector-valued stochastic process describable by a mathematically tractable stochastic linear recursion relation. Let

$$\vec{\xi} : N_0 \longrightarrow R^N \tag{24}$$

(where $\vec{\xi}(t)$ is the state at time $t \geq 0$) denote the process, and

$$\vec{\xi}(0) = \vec{c} \in R^N \tag{25}$$

the initial state. Then the evolution equation becomes

$$\vec{\xi}(t) = A\vec{\xi}(t-1) + \vec{\Gamma}(t) \ , \quad t > 0 \tag{26}$$

with $A$ the relaxation matrix

$$A = \begin{pmatrix} 1-\frac{1}{P} & \frac{1}{P} & 0 & \cdots & & 0 \\ \frac{1}{P} & 1-\frac{2}{P} & \ddots & \ddots & & \vdots \\ 0 & \ddots & \ddots & \ddots & & 0 \\ \vdots & \ddots & \ddots & 1-\frac{2}{P} & \frac{1}{P} \\ 0 & \cdots & 0 & \frac{1}{P} & 1-\frac{1}{P} \end{pmatrix} \quad (N \times N - \text{matrix}) \tag{27}$$

and $\vec{\Gamma}(t)$ the stochastic perturbation. This relaxation differs from the one considered in section 3.2 through the boundary condition for the first and last site: The relaxation matrix here ensures conservation of the overall sum of slopes in the vector $\vec{\xi}$. The random vector $\vec{\Gamma}(t)$ has $N-1$ equally probable outcomes,

$$\vec{\Gamma}^{(i)} := (0, \ldots, 0, -s, \underbrace{s}_{i.\text{th comp.}}, 0, \ldots, 0) \ , \quad i = 2, \ldots, N \ . \tag{28}$$

For different times the perturbation vectors are independent random vectors.

The matrix $A$ has eigenvalues

$$\lambda_k = 1 - \frac{2}{P}\left(1 - \cos\frac{(k-1)\pi}{N}\right) \ , \quad k = 1, \ldots, N \tag{29}$$

with corresponding eigenvectors $\hat{\vec{v}}^{(1)}, \ldots, \hat{\vec{v}}^{(N)}$,

$$\left.\begin{array}{rcl} \hat{v}_i^{(1)} &=& \frac{1}{\sqrt{N}} \ , \quad i = 1, \ldots, N \\ \hat{v}_i^{(k)} &=& \sqrt{\frac{2}{N}}\cos\frac{(k-1)\pi(2i-1)}{2N} \ , \\ & & k = 2, \ldots, N \ , \ i = 1, \ldots, N \end{array}\right\} \tag{30}$$

which we combine in the matrix

$$W := (w_{ij})_{i,j=1}^{N} := (\hat{v}_i^{(j)})_{i,j=1}^{N} \quad . \tag{31}$$

Note that $\lambda_1 = 1$ and $|\lambda_k| < 1$ for $k > 1$, $P > 2$.

In the following we want to derive out of the known statistical properties of the perturbation process $\vec{\Gamma}$ the statistical properties of the process $\vec{\xi}$. In particular, we will calculate the time correlation function and the spectral density function. The process $\vec{\Gamma}$ has the averages

$$\left. \begin{array}{rl} < (\Gamma(t))_1 > & = -\frac{s}{N-1} \\ \\ < (\Gamma(t))_i > & = 0 \quad \text{for} \ \ i = 2, \ldots, N-1 \\ \\ < (\Gamma(t))_N > & = \frac{s}{N-1} \end{array} \right\} \tag{32}$$

and the variances

$$\begin{aligned} g_{ij} &:= \ < [(\Gamma(t))_i - < (\Gamma(t))_i >] [(\Gamma(t))_j - < (\Gamma(t))_j >] > \\ &= \ < (\Gamma(t))_i (\Gamma(t))_j > \ - \ < (\Gamma(t))_i > < (\Gamma(t))_j > \\ &= \ \begin{cases} \frac{s^2}{N-1} - \frac{s^2}{(N-1)^2} & , \ i = j = 1 \ \text{or} \ i = j = N \\ \\ \frac{2s^2}{N-1} & , \ i = j \neq 1, N \\ \\ -\frac{s^2}{N-1} & , \ |i - j| = 1 \\ \\ \frac{s^2}{(N-1)^2} & , \ i = 1, j = N \ \text{or} \ i = N, j = 1 \\ \\ 0 & , \ \text{else} \end{cases} \end{aligned} \quad . \tag{34}$$

A central role in the derivation is played by the formal solution of the recursion relation (26):

$$\forall \ t \geq 0 \ : \ \vec{\xi}(t) = \sum_{m=1}^{t} A^{t-m} \vec{\Gamma}(m) + A^t \vec{c} \quad . \tag{35}$$

The powers of $A$ can be expressed in the form

$$(A^\tau)_{ij} = \sum_{k=1}^{N} W_{ik} W_{jk} \lambda_k^\tau \quad . \tag{36}$$

Insertion of (36) in (35) and calculation of the expectation value leads to

$$< \xi_i(t) > \ = \ -\frac{s}{N-1} \sum_{m=1}^{t} \sum_{k=2}^{N} W_{ik} W_{1k} \lambda_k^{t-m}$$

$$+ \frac{s}{N-1} \sum_{m=1}^{t} \sum_{k=2}^{N} W_{ik} W_{Nk} \lambda_k^{t-m}$$

$$+ \sum_{j,k=1}^{N} W_{ik} W_{jk} \lambda_k^t c_j \quad . \tag{37}$$

Since we are interested in the stationary state we take the limit $t \to \infty$:

$$\begin{aligned} <\xi_i(\infty)> &:= \lim_{t \to \infty} <\xi_i(t)> \\ &= \frac{Ps}{N-1} i - \frac{Ps}{2} \frac{N+1}{N-1} + \frac{1}{N} \sum_{j=1}^{N} c_j \quad . \end{aligned} \tag{38}$$

A similar calculation can be done for the second central moments of the random vector $\vec{\xi}(t)$:

$$\sigma_{ij}(t) := <(\xi_i(t) - <\xi_i(t)>)(\xi_j(t) - <\xi_j(t)>)> \quad . \tag{39}$$

In the stationary state we find

$$\sigma_{ij}(\infty) := \lim_{t \to \infty} \sigma_{ij}(t) \tag{40}$$

$$= \sum_{r_1, r_2=2}^{N} \sum_{k_1, k_2=1}^{N} g_{k_1 k_2} W_{ir_1} W_{k_1 r_1} W_{jr_2} W_{k_2 r_2} \frac{1}{1 - \lambda_{r_1} \lambda_{r_2}} \quad . \tag{41}$$

The temporal correlation functions of the process $\vec{\xi}$ are defined by

$$K_{ij}(\tau, t) := <(\xi_i(t+\tau) - <\xi_i(t+\tau)>)(\xi_j(t) - <\xi_j(t)>)> \tag{42}$$

for $t \geq 0$, $\tau \in Z$ such that $t + \tau \geq 0$. In the stationary state they can be calculated to be

$$\begin{aligned} \forall \tau \geq 0 \; : \; K_{ij}(\tau) &:= \lim_{t \to \infty} K_{ij}(\tau, t) \\ &= \sum_{r=2}^{N} \sum_{k=1}^{N} W_{ir} W_{kr} \lambda_r^\tau \sigma_{kj}(\infty) \end{aligned} \tag{43}$$

$$\forall \tau \in Z \; : \; K_{ij}(\tau) = K_{ji}(-\tau) \tag{44}$$

The spectral density function is the Fourier transform of the autocovariance function,

$$\begin{aligned} S_{jj}(\omega) &:= \frac{1}{2\pi} \sum_{\tau \in Z} K_{jj}(\tau) e^{-i\omega\tau} \quad , \omega \in [-\pi, \pi] \\ &= \frac{1}{2\pi} \left( \sigma_{jj}(\infty) + \sum_{k=1}^{N} \sum_{r=2}^{N} W_{jr} W_{kr} \sigma_{kj}(\infty) \frac{2\lambda_r \cos\omega - 2\lambda_r^2}{1 - 2\lambda_r \cos\omega + \lambda_r^2} \right) \quad . \end{aligned} \tag{45}$$

The autocovariance function and spectral density function calculated from (43) and (45), respectively, are compared with simulation results in Fig. 9. The simulated data cannot be extended to very low frequencies because of time and memory constraints on the time series analysis. The calculated spectral density function shows the two different regions of algebraic behaviour as mentioned in section 2.3. For very low frequencies it approaches the value $S_{jj}(0)$.

The mechanism for the short time correlations which correspond to the higher frequencies is the damping of perturbations. For $P > 4$ all eigenvalues of the relaxation matrix are positive. The one with the smallest value

$$\lambda_N = 1 - \frac{2}{P}\left(1 - \cos\frac{(N-1)\pi}{N}\right) \cong 1 - \frac{4}{P} \tag{46}$$

corresponds to the fastest decay and may thus serve as a measure for the relaxation rate. An excitation of this eigenvalue will decay as $(1-4/P)^t$. The typical relaxation time $T$ can then be estimated from $(1 - 4/P)^T = 0.01$ (decay to 1%) or $T = \ln 0.01/\ln(1 - 4/P)$. This time corresponds to the frequency

$$\omega = \frac{\pi}{T} = \frac{\pi}{\ln 0.01}\ln(1 - \frac{4}{P}) \cong \frac{\pi}{\ln 0.01}\left(-\frac{4}{P}\right) \cong \frac{\pi}{P} \tag{47}$$

which is quite a good estimate of the transition frequency between the two scaling regions. For instance for the parameter choice used in Fig. 9 this estimate predicts the second transition frequency to be near $\log \omega = -0.5$.

We suspect perturbations at different time steps to be the mechanism responsible for the long time correlations which correspond to low frequencies. That the low frequency cut off in the spectral density function is due to a finite size effect can be seen through inspection of eq. (45): $S_{jj}$ is a superposition of functions of the form

$$f_\lambda(\omega) = \frac{2\lambda \cos\omega - 2\lambda^2}{1 - 2\lambda\cos\omega + \lambda^2} \quad , \quad \lambda \in ]0,1[. \tag{48}$$

For small frequencies this is a Lorentzian

$$f_\lambda(\omega) = \frac{2\lambda(1-\lambda) - \lambda\omega^2}{(1-\lambda)^2 + \lambda\omega^2} \quad . \tag{49}$$

The Lorentzian in eq. (45) with the highest maximum and the narrowest width is the one corresponding to the eigenvalue closest to 1, i.e.

$$\lambda_2 = 1 - \frac{2}{P}(1 - \cos\frac{\pi}{N}) \cong 1 - \frac{1}{P}\frac{\pi^2}{N^2} \quad . \tag{50}$$

The transition frequency $\omega_c$ may be estimated by that frequency where the dominant Lorentzian drops to a fraction $c$ of its maximum, i.e.

$$f_{\lambda_2}(\omega_c) = c f_{\lambda_2}(0) \quad , \quad c \in ]0,1[ \quad . \tag{51}$$

After a few steps one arrives at:

$$\omega_c = \sqrt{\frac{2(1-\lambda)^2(1-c)}{(1-\lambda) + 2c\lambda}} \cong (1-\lambda)\sqrt{\frac{1-c}{c}}$$

$$\cong \frac{1}{P}\frac{\pi^2}{N^2}\sqrt{\frac{1-c}{c}} \tag{52}$$

The choice $c = 0.95$ leads to an estimate for the low frequency cut off in Fig. 9 of $\log \omega = -3.4$. A plot of eq. 52 for increasing lattice size $N$ shows a decrease in the estimated transition frequency which supports the assumption of a finite size effect.

## 5 Summary and conclusions

In this paper we have analyzed a stochastically driven running one-dimensional sandpile automaton by looking at the dynamics on a finite lattice. We could explain qualitatively the stationary slope profiles and calculate the transition sites between the different phases. The profiles are similar to the ones of the limited local sandpile model found in [3]. The limited local model lacks a linearly increasing part since it does not incorporate weighted relaxation.

The kind of transport argument used here is well adapted to discrete time dynamics on a finite lattice and allows to characterize the stationary state without looking at a finite size scaling theory for the corresponding continuum model. It also helps to avoid problems met when approximating the dynamics by a continuous, nonlinear diffusion equation as encountered in [3]. In particular, the ability to achieve the required transport via undamped running waves seems to be the origin of the singularity of the diffusion constant in such models.

We calculated exactly the nontrivial scaling autocovariance functions and spectral density functions of time series recorded in phase III using well known methods for stochastically perturbed linear recursion relations. The final form for the spectral density function, eq. (45), allows to estimate the transition frequencies between the scaling regions. It also contains the scaling behaviour but we have not succeeded in estimating exponents analytically.

How much of these observations and methods generalizes to higher dimensional automata is not obvious. One might expect to find spherical waves instead of our linear waves, but as the wave propagates outwards the mass it transports is diluted up to the point where it stops propagating. This transition between running and stationary sites complicates the analysis. Similarly, because of the dilution it is difficult to build up the linear, always supercritical profile, in which one could apply methods for stochastically driven linear maps. The easiest way to achieve this is an 'inverted sandpile', where sand is drained in the center and not at the circumference of the pile. Since the mass is then transported towards the center, the flux rates increase and the sites will always be supercritical. This might serve as a model for the formation of a hydrodynamic drainage network (see for instance the experiment described in [6]).

# Acknowledgement

We thank H.J.Schellnhuber for support and discussions during an early stage of this work.

Figure 1: Schematic picture of the linear automaton before (a) and after (b) a perturbation. $h_i$ denotes the height at site $i$ and $\sigma_c$ the slopes. The values after a perturbation are marked by *.

Figure 2: Division of the $P-s-$plane. The crosses mark the parameter values used in the simulation and are accompanied by references to figures which contain the stationary slope profiles. The two continous lines divide the plane into three different parts characterized by distinct structures of the profiles (see the text for details). They are predictions which are obtained by transport arguments described in subsection 3.3. The two dotted lines show the divergence points of the curves. For parameters chosen out of the shaded region the automaton does not converge to a stationary state.

Figure 3: Expectation value of the slope profile for the parameter choice $P=6$, $s=0.25$ and snapshots of the time evolution. The profile is a nearly constant subcritical line. Subcritical wave packets are characteristics of the dynamics in all parts of the lattice.

Figure 4: Expectation value of the slope profile for the parameter choice $P=6$, $s=1.0$ and snapshots of the time evolution. Compared to Fig. 3 one observes a second regime in the slope profile: A regime with supercritical slope in which the dynamic is dominated by supercritical wave packets. The two parts of the profile are nearly horizontal.

Figure 5: Expectation value of the slope profile for the parameter choice $P=6$, $s=2.5$ and snapshots of the time evolution. Compared to Fig. 4 this profile shows more structure: There is an additional third part which increases linearly and in which perturbations are damped.

Figure 6: Expectation value of the slope profile for the parameter choice $P=1.5$, $s=1.0$ and snapshots of the time evolution. This pair of parameters lies in the same region of the phase diagram in Fig. 2 as the one used in Fig. 3. The only difference is a slight ascent in the last part of the lattice.

Figure 7: Expectation value of the slope profile for the parameter choice $P=1.5$, $s=2.5$ and snapshots of the time evolution. These parameters are taken out of region II of the phase diagram in Fig. 2; one has to compare this profile with the one given in Fig. 4. The dynamics observed in sub- and supercritical parts agrees, while the plateaus in Fig. 4 degrade to a monotonically increasing profile.

Figure 8: Autocovariance function (a) and spectral density (b) for a time series recorded at site 160 in the automaton with parameters $P=10$, $s=2$. The site 160 lies in the linearly increasing part of the stationary profile. The two scaling regions in the spectral density have slopes -0.6 and -1.44, respectively.

Figure 9: Measured and calculated (from (43) and (45)) autocovariance function (a) and spectral density function (b) of the slope at site 25 in the model described by the recursion relation (26) with parameters $P = 10$, $s = 2$, $N = 50$. The calculated curves are shifted by 0.5 units of the vertical axis. The two scaling regions in the calculated curve have scaling exponents of -0.53 and -1.45, respectively.